# Socially-Aware Networking: A Survey

Feng Xia, *Senior Member, IEEE*, Li Liu, Jie Li, Jianhua Ma, and Athanasios V. Vasilakos, *Senior Member, IEEE*

*Abstract*—The widespread proliferation of handheld devices enables mobile carriers to be connected at anytime and anywhere. Meanwhile, the mobility patterns of mobile devices strongly depend on the users' movements, which are closely related to their social relationships and behaviors. Consequently, today's mobile networks are becoming increasingly human centric. This leads to the emergence of a new field which we call socially-aware networking (SAN). One of the major features of SAN is that social awareness becomes indispensable information for the design of networking solutions. This emerging paradigm is applicable to various types of networks (e.g. opportunistic networks, mobile social networks, delay tolerant networks, ad hoc networks, etc) where the users have social relationships and interactions. By exploiting social properties of nodes, SAN can provide better networking support to innovative applications and services. In addition, it facilitates the convergence of human society and cyber physical systems. In this paper, for the first time, to the best of our knowledge, we present a survey of this emerging field. Basic concepts of SAN are introduced. We intend to generalize the widely-used social properties in this regard. The state-of-the-art research on SAN is reviewed with focus on three aspects: routing and forwarding, incentive mechanisms and data dissemination. Some important open issues with respect to mobile social sensing and learning, privacy, node selfishness and scalability are discussed.

*Index Terms*—Social Awareness; Mobile Social Networks; Data Dissemination; Routing; Forwarding

## I. INTRODUCTION

THE last few decades have witnessed striking developments in wireless communications and networking technologies which yield essential network architectures to constitute ubiquitous networks. The rich diversity of wireless mobile devices, especially smart mobile devices, has joined in the networks with unprecedented speed. As predicted by Gartner, mobile phones will overtake PCs as the most common Web access devices in the world by 2013 [1]. Clearly, the applications and services are being migrated into mobile devices. Thus, mobility, or more generally the dynamic nature stands out as the main factor affecting the service. Many applications benefit from the mobile devices, such as mobile social network services, environment monitoring, urban sensing, etc. Currently, a great number of mobile devices access the Internet through e.g. 3G cellular networks, which leads to huge communication overload and thus has negative influence on the quality of service (QoS) of the applications.

With the increasing workload, it has become crucial to make full use of the limited resources of mobile devices and communication networks so that the resource efficiency can be improved and hence more and more mobile applications can be supported. Recently, researchers begin to address this issue from a social perspective. As a consequence, the field of socially-aware networking (SAN) is emerging as a new paradigm to exploit the social properties of network nodes (especially mobile devices) for designing networking solutions. In addition, recent rapid advances in social networking applications are also a major drive force for the emergence of SAN.

The main starting point of SAN relies on the dramatic development on fields of wireless communications and sociology (or social network theory). Firstly, the mobile devices are now equipped with more storage, higher computing speed and various wireless communication interfaces. By using short-range wireless technologies such as Wi-Fi and Bluetooth, mobile devices could form temporary ad hoc networks to communicate with each other. In recent years, many research efforts related to mobile devices have been made in order to explore the potential of mobile devices. For instance, opportunistic networks [2] and mobile ad hoc networks (MANET) [3] aim at exploring communications among mobile devices by using the aforementioned short-range telecommunication protocols. On the other hand, mobile sensing [4], mobile cloud computing [5] and opportunistic computing [6] utilize the huge information resources and computing capability provided by mobile devices. However, ad hoc networks are not stable enough which usually utilize the opportunistic contact to communicate. The mobility of mobile devices will directly affect the topology of networks.

Secondly, it has been found that mobile devices have close relationship with objects in society, because they are usually carried by e.g. people, animals or vehicles. The mobility of devices' carriers (i.e. users) represents the mobility of mobile devices. Human beings, as the main actors of mobility, take especially important role and attract increasing attention from information technology both in academy and industry. There are a lot of social properties or relationships hidden

This work was partially supported by the Natural Science Foundation of China under Grant No. 60903153, Liaoning Provincial Natural Science Foundation of China under Grant No. 201202032, and the Fundamental Research Funds for the Central Universities (DUT12JR10).

Feng Xia, Li Liu, and Jie Li are with School of Software, Dalian University of Technology, Dalian 116620, China (e-mail: f.xia@ieee.org).

Jianhua Ma is with the Faculty of Computer and Information Sciences, Hosei University, Japan (e-mail: jianhua@hosei.ac.jp).

Athanasios V. Vasilakos is with the Dept of Computer and Telecommunications Engineering, University of Western Macedonia, Greece (e-mail: vasilako@ath.forthnet.gr).

among human beings. For example, people are often highly sociable and people with similar properties usually spend long time together, being more willing to share information and resources. The gregarious feature is called community. People from the same community may contact or share information with higher probability. By exploring social relationships, catching the mobility regularities of mobile devices as well as predicting the contact opportunities of them will be effective. Furthermore, the contact information could be a basic evidence for designing the routing protocols in mobile environments.

SAN has built its own features driven by the aforementioned spheres. First of all, SAN focuses on wireless communications among ad hoc networks which generally consist of mobile devices connected via e.g. Wi-Fi, Bluetooth, etc. Similar to opportunistic networks, SAN is characterized by the intermittent disconnection and the absence of the infrastructure. Furthermore, social properties are valuable to find the users' mobility pattern and predict contact opportunities more accurately. Therefore SAN is able to discover more reliable contact opportunities (with higher prediction accuracy) by taking advantage of the social awareness of network nodes. Social awareness (also known as social consciousness) originates from sociology, and it is often used to describe the sociability and social behaviors of human beings. For instance, social awareness could be consciously shared within a society, which means that a user should know what is socially acceptable and what to be performed in that manner.

The SAN paradigm can be applied into many areas such as pocket switched networks (PSN) [7], vehicular ad-hoc networks (VANET) [8] and cyber-physical systems (CPS) [9], etc. In the meantime, it requires support from a number of technologies such as mobile sensing, opportunistic computing, and social network analysis. Recently, several research fields have emerged as socially aware paradigms which share some common attributes with SAN, such as mobile social networks [10], social opportunistic networks [11], vehicle social networks [12], opportunistic Internet of Things (IoT) [13], etc. They all address mobile nodes and are driven by social network theory. However, they are different in terms of focus. Mobile social networks originate from online social networks which emphasize the relationship of similar interests or commonalities. Social opportunistic networks are based mainly on the prediction of contact opportunities and concentrate on the social relationships between mobile nodes and their regular movements. Vehicle social networks consider the social properties of vehicles and passengers. Opportunistic IoT usually focuses on the interaction between human and IoT.

Compared to these related areas, SAN is human-centric and studies individuals' social properties comprehensively. They involve a great deal of information, such as personal property (e.g. preference, habit, and life regularity), human-to-human relationship (e.g. friendship, colleague, and family), human-to-community relationship (e.g. interests and popularity), and human-to-environment relationship. Based on these social properties, the objective of SAN is to design networking solutions to support mobile applications. In this context, routing and forwarding protocols and dissemination strategies are the most important components by which data is delivered or disseminated. In addition, mobile devices usually involve personal (or private) information which leads to selfish behaviors that benefit the users. Thus incentive mechanisms are also an essential component.

In this paper, we present a survey of the emerging field of socially-aware networking. To the best of our knowledge, this is the first effort of its kind. We present the architecture of SAN. Following a discussion on social properties, we summarize the state-of-the-art research efforts on routing and forwarding protocols, incentive mechanisms and data dissemination algorithms in SAN. Our aim is to provide a better understanding of research opportunities and challenges in the field of SAN and to find appealing hints for future explorative activities on this timely and exciting topic.

The remainder of this paper is structured as follows. Section II presents the architecture of SAN and Section III describes several important social properties. In Section IV, we review socially-aware routing and forwarding protocols from perspectives of unicast and multicast, respectively, along with some discussion on congestion problem. Incentive mechanisms for dealing with node selfishness are explored in Section V. The paper examines socially-aware data dissemination algorithms categorized into solicitation and cache based approaches and forwarding based approaches in Section VI. We discuss some important open issues in Section VII and conclude the paper in Section VIII.

## II. Architecture of Socially-Aware Networking

SAN studies the context information of network nodes (especially mobile devices), captures and extracts the social properties and formulates reasonable protocols to support upper-layer applications. However, SAN is an emerging field still in its infancy. Till now there is no formal (uniform) definition for this terminology or its architecture. In this paper, we propose a simple architecture of SAN, as shown in Fig. 1 where the above mentioned workflow of SAN is depicted. The detailed description about the structure is presented below.

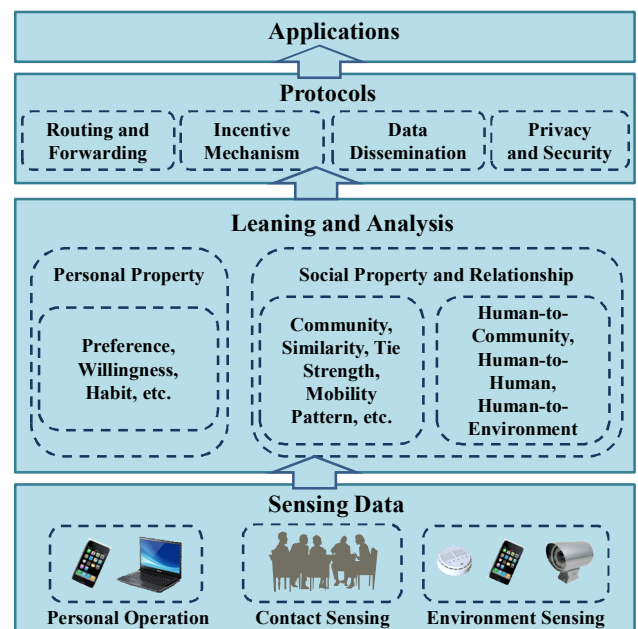

Fig. 1. Architecture of SAN

The major task of the first two steps is to achieve social awareness by sensing and analyzing data using some intelligent technologies, such as data mining and machine learning. To obtain the information about personal behaviors, social contact and the environment, mobile sensing is an effective way to sense the surroundings. Specifically, various sensors deployed in real world could be utilized, as well as sensors embedded in mobile devices. The available sensors include e.g. accelerometer, digital compass, gyroscope, GPS, microphone and camera, which enable personal, group, and community scale sensing applications [4]. In this sense, SAN relies on CPS, i.e. to obtain the information of social awareness. On the other hand, SAN enhances CPS by exploiting social properties, which enrich human-to-human, human-to-object, and object-to-object interactions in the physical world as well as in the virtual world.

The various data, which reflect either personal behaviors or environment information, are the main evidences to explore both the personal and social properties. The personal properties involve e.g. one's preference, willingness, habit, life regularity, etc. They could be obtained through analyzing the operations and behaviors of the user. SAN is based on social properties and relationships of users, which are more reliable and more natural. The learning and deduction process requires more resources and computing capability, which might need the support of e.g. mobile cloud computing and data mining.

Through analyzing and learning from these data, SAN can deduce important social properties. The most commonly used social properties include community, centrality, similarity, tie strength and human mobility pattern. These properties are connected to the social relationships. To be exact, the human-to-human relationship refers to direct contact relationship and it involves personal information such as friendship, which indicates a kind of tie strength. The human-to-community relationship indicates an individual's gregarious property and status in community. They can be gathered by use of contact graph (physical contact) and social graph (virtual graph, such as online social networks). The human-to-environment relationship implies that the environmental information is related to human beings.

The available personal and social properties are valuable evidence to devise the routing protocols and data dissemination algorithms. Thus, socially-aware routing and forwarding protocols and socially-aware data dissemination algorithms are essential issues in SAN. In addition, the protocols in SAN rely heavily on the nodes' cooperation due to the exploration of social relationship and interactions among the users. As a consequence, incentive mechanisms are necessary to deal with selfishness of nodes. In addition, privacy and security protecting strategies are also an important component of SAN. However, it is very challenging and we list it in open issues. For these reasons, routing and forwarding protocols, incentive mechanisms and data dissemination algorithms in the context of SAN are the main topics of this paper. Fig. 2 illustrates a generalized taxonomy for SAN protocols based on the above description, which is used for organizing this paper.

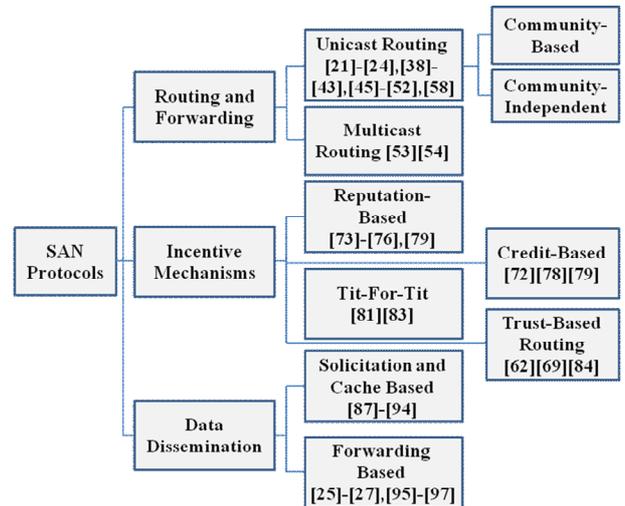

Fig. 2. Taxonomy of Protocols in SAN

SAN has some similarities with the SCI (social and community intelligence) concept presented by Zhang *et al.* [14]. However, SCI focuses on the collection and analysis of social information and aims to reveal individual and group behaviors, social interactions and community dynamics, while SAN pays more attention to designing networking protocols by use of social properties. It is worth noting that the proposed architecture is just a starting point for extensive study of SAN. We hope it will eventually lead to a converged viewpoint and taxonomy of SAN.

## III. SOCIAL PROPERTIES

The research on social networks was initiated by Milgram in 1960's [15]. Milgram introduced the idea of small world phenomenon which indicates that any pair of people in the world can connect with each other through small sequences of relationships (typically five or six), and thereafter many works reaffirmed it [16]-[18].

Contemporary researches mainly focus on understanding the dynamics and structure of social networks with respect to relationships which can be classified in terms of strength of mutual familiarity and intensity. Social network analysis (SNA) [19] studies relationships between human beings, as well as patterns and implications of these relationships. SNA is a useful and powerful tool for analyzing complex social relationships among people in social sciences [20]. The notion of social network and the methods for SNA attract significant interest initially from the social and behavioral communities, later data mining, and only recently from the networking community. In this section, we concentrate on social properties that are most popular in design of protocols and algorithms in SAN.

*A. Social Graph*

Social networks exhibit the small world phenomenon that node encounters are sufficient to build a connected relationship graph. Graph is a convenient tool to represent the relational structure of social networks in a natural manner, which is generally called social graph. In a social graph, vertices (nodes) indicate human individuals and edges (links) indicate social relationships between individuals. In some degree, social network is equal to the social graph, so they can be used alternatively. One significant challenge in social networks is how to represent a link between two nodes [21]. According to different link meanings, several social graphs are proposed in recent literature. Contact graph is a popular way to analyze and estimate relationships among people by observing their inter-contact time in the history [22]-[25]. Besides, neighbor graph [21], regularity graph and interest graph [27] are proposed recently as well.

*B. Community*

A community is a structural subunit (which can be represented as a set of individuals) of a social network with high density of internal links [21]. Individuals have more social connections with other individuals inside their own community than with individuals outside. The social connections may be family, friends, common location or common interest, which are decided by the social graph. In general, individuals in the same community may meet each other more frequently. Therefore, community structure has significant impact on people's mobility patterns and thus is beneficial for choosing appropriate forwarding path.

*C. Centrality*

Besides community, the centrality is another basic concept in social networks, which considerably affects the performance of socially-aware forwarding algorithms. The experiments in [28] demonstrate that it is important to find appropriate centrality and community in the design of socially-aware forwarding and data dissemination algorithms.

Centrality is used to describe important and prominent nodes in a social graph. People have various roles and popularities in society. A central node has stronger capability of connection with other nodes. Experiments in [24] show that there is a small number of nodes which have extremely high relaying ability, and a large number of nodes which have moderate or low centrality values, with 30 and 70 percentiles, respectively. This phenomenon is called *human heterogeneity* (or *node heterogeneity*).

The most recognized centrality measures are introduced by Freeman [29][30]: degree centrality, betweenness centrality and closeness centrality. Here we give a brief introduction to each of them.

*1) Degree Centrality*

Degree centrality is defined as the number of one-hop neighbors of a node [19]. For a network consisting of *n* nodes, the degree centrality $DegC_i$ of a node *i* is:

$$DegC_i = \frac{deg(i)}{n-1} \quad (1)$$

where *deg(i)* is the number of directly connected neighbors of node *i*. $DegC_i$ indicates the connection ratio of node *i* between the real connection number of node *i* and the maximum number of possible connections, i.e. *n*-1. Degree centrality identifies the most active nodes in the network. A node with high degree centrality maintains large number of links to others. As such, a central node occupies a structural position (network location) that may act as a conduit for information exchange [23].

*2) Betweenness Centrality*

Betweenness centrality is the percentage of the number of shortest-paths including node *i* over all the shortest-paths [31]. The betweenness centrality $BetC_i$ of a node *i* is:

$$BetC_i = \sum_{j \neq i} \frac{sp_{j,k}(i)}{sp_{j,k}} \quad (2)$$

where $sp_{j,k}$ is the number of shortest paths linking nodes *j* and *k*, and $sp_{j,k}(i)$ is the number of shortest paths linking nodes *j* and *k* which pass through node *i*. Betweenness centrality is a measure of the extent to which a node has control over information flowing between others [32]. Nodes with high betweenness centrality can bridge two nonadjacent nodes and may impact on data flow between communities. Therefore, betweenness centrality is a key metric to determine the links between communities.

*3) Closeness Centrality*

Closeness centrality is defined as the inverse of the sum of the distances between a given actor and all other actors in the network [14]. The closeness centrality $CloC_i$ of a node *i* is

$$CloC_i = \frac{1}{\sum_{i \neq j} dis(i,j)} \quad (3)$$

where *dis(i,j)* is the distance between nodes *i* and *j*. A node with the highest closeness means that the node has the shortest path to other nodes in network. So closeness centrality describes the efficiency of information propagation from one node to all the others. In message forwarding and data dissemination applications, closeness centrality can be used to choose relay nodes to deliver the message with success and/or faster to the other nodes in the community.

The analysis of data collected from ACM CoNEXT'07 concludes that centrality is the primary factor to decide whether a node is a good next hop and the best performance trade-off is obtained when several complementary rules are combined [33].

*D. Similarity*

Sociologists have long realized that social network displays a high degree of transitivity, that is to say, there is a heightened probability of two people being acquainted if they have one or more other acquaintances in common. This phenomenon is also called *clustering* [23]. Similarity indicates the group of nodes depending upon common contacts or interests which can be measured by the ratio of common links (e.g. contact, interest, neighbors) between individuals. The higher similarity a node and the destination share, the more opportunities they have to encounter. Nodes with higher similarity can be good candidates for information dissemination among clusters of nodes.

*E. Tie Strength*

The notion of tie strength was firstly introduced by Granovetter in 1973, which is defined as "the amount of time,

the emotional intensity, the intimacy (mutual confiding), and the reciprocal services, which characterize a tie" [34]. Tie strength is a quantifiable property that characterizes the link between two nodes. Strong ties are more likely to be activated for information flow when compared to weak ties. The most widely used tie strength indicators are: frequency, intimacy/closeness, longevity, reciprocity, recency, multiple social context, and mutual confiding (trust) [22]. A combination of the tie strength indicators can be used for information flow to determine which contact has the strongest social relationship to the destination. Meanwhile, the effects of weak ties in social networks are also crucial to data dissemination. Granovetter's research modifies that the weak ties may be beneficial to forming bridges between high density clusters.

### F. Human Mobility Pattern

Researchers observe that human mobility presents two key properties from analysis of real mobility traces: *spatial regularity* and *temporal regularity* [35][36]. Spatial regularity is that nodes usually move around a set of locations frequently and regularly in time schedule. For example, the students usually move around among dormitory, classroom and canteen at different periods of the day. Temporal regularity is that human mobility pattern is repetitive at long time. For instance, human repeats working mobility pattern from Monday to Friday. The two properties of human mobility can be used to predict the users' future mobility which plays important role in choosing the forwarders.

## IV. SOCIALLY-AWARE ROUTING AND FORWARDING PROTOCOLS

Due to the mobility of mobile devices, it is usually difficult to find an end-to-end path between source and destination(s) at the beginning of communications. All of the current routing methods share a store-carry-and-forward paradigm and utilize contact opportunity to communicate in intermittently (dis)connected networks such as delay tolerant networks (DTN) and opportunistic networks. Thus relaying selection and forwarding decision are critical to be made by the current node based on certain routing strategy. The prediction of contact opportunity is one key issue for design of efficient routing and forwarding protocols. SAN considers the social properties during the course of routing and forwarding protocol design to make better forwarding decisions.

Routing can be classified into unicast and multicast according to the number of destinations the data need to be delivered. Unicast routing focuses on forwarding data to a single specific destination. Multicast involves the distribution of the data to a group of users. In this section, we will review unicast routing protocols and multicast routing protocols based on social properties, respectively. In addition, congestion is a very challenging problem as well as an important factor that influences the design of routing protocols. Consequently, here we will also examine state-of-the-art work addressing the congestion problem in the context of SAN.

### A. Unicast Routing

The key faction of routing protocols is to select the most optimal relay nodes which have highest probability to meet the destination(s) in order to maximize message delivery ratio while minimizing message overhead and delay. The relative stable social relationships (especially the community relationship) of the users are reasonable information for predicting future contact opportunities. The social network structure can be viewed as three levels: individual, community, and whole network [37]. The community is very widely used in routing protocols. Nodes in the same community have more chances to meet than those in different communities. A large number of community detection algorithms have been proposed to divide a social network into separated communities. Based on whether the support of community is needed or not, we classify routing and forwarding protocols as community-based routing and community-independent routing.

#### 1) Community-Based Routing

It is believed that nodes have more opportunities to contact in community, which is beneficial to forwarding messages for other members of their community. One of the first research works on community in routing protocols is carried out by Hui and Crowcroft [38]. They conducted an experiment in INFOCOM'06 conference, which results in community relationship through analyzing inter-contact time distribution. The results proved that community structure can improve forwarding efficiency. *Inter-contact time* is the time elapsed between two successive contact periods for a given pair of devices [39]. In experimental analysis of inter-contact time distribution, the intra-community pair has higher power law coefficient than the inter-community pair, which indicates that nodes in the same community tend to meet more often. In the experiment, they proposed a forwarding scheme named LABEL which assumes each node has a label on behalf of its affiliation. It directly forwards messages to destination, or chooses next-hop nodes with the same label as the destination node. They also presented the concept of friendship community which can help improve delivery.

The general operation principle in community-based routing is shown in Fig. 3. Community is the basis of forwarding data. Firstly, mobile nodes are grouped into communities by certain community detection algorithm. According to the forwarding strategy, data are forwarded among nodes. If the relay nodes are out of destination community (the community where the destination belongs to), the inter-community forwarding strategy is used to forward data close to the destination community as quickly as possible. If the relay nodes are in the destination community, the intra-community forwarding strategy is used to forward data to the central node and the central node can forward the data to the destination.

Community detection and formation play an important role in community-based routing. In LABEL, the affiliation (or community) is labeled through configuration. To make the community practically useful, many community detection algorithms have been proposed. Most of the proposed algorithms are centralized and they focus on analysis of offline mobile trace. Contrary to the centralized methods (see [39]), Hui and Crowcroft proposed three algorithms, named SIMPLE, *K*-CLIQUE and MODULARITY, for online distributed community detection based on the contact graph in [24]. Then they presented BUBBLE RAP [24] based on socially-aware overlay. The overlay is constructed by combining the community with centrality. BUBBLE RAP calculates global ranking (i.e. global centrality) for the whole network and local ranking (i.e. local centrality) for the local community for each node based on degree centrality. When a node has a message destined for another node, the node first bubbles the message up through hierarchical ranking tree using the global ranking until reaching a node which is in the same community as the destination node. Then in the same community, the message is bubbled up through local ranking tree using the local ranking until the destination is reached or message expires.

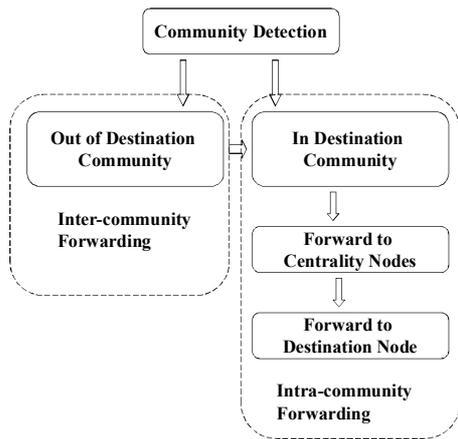

Fig. 3. Operating Principle of Community-based Routing

Bulut and Szymansk [41] introduced direct and indirect friendship to indicate the tie strength of virtual link and to form friendship community. They consider three behavior features of close friendship: high frequency, longevity and regularity, and also define two metrics called SPM (social pressure metric) and CSPM (conditional SPM) for direct and indirect friendship based on contact history. Friendship community is a set of nodes having a link quality larger than a threshold. To reflect temporal distinctions, different friendship communities in different periods of the day are established for each node. When forwarding, current node with the message will choose the node that belongs to the same community with the destination and with a stronger friendship of destination than current node.

Several works in the literature consider the relationship between communities and formulate inter-community forwarding decision [21][42][43].

LocalCom [21] only requires limited local information to form communities and it also considers the forwarding between different communities. Firstly, the authors presented a metric named "*similarity*" to construct the neighboring graph which considers the encounter frequency, encounter length, separation period in the encounter history. Then based on the neighboring graph and the associated "*similarity*", the authors proposed a distributed algorithm for community detection, and represented the communities with an extended *k*-hop clique. LocalCom adapts different forwarding plans for intra- and inter-community packet forwarding. Intra-community packet forwarding adapts single hop source routing. The "*similarity*" metrics indicate the quality of virtual links (tie strength) and the node with larger "*similarity*" metrics is chosen to forward data. Controlled flooding routing is used for inter-community communication based on *gateways*. *Gateways* are nodes that have direct neighboring relationships with nodes in other communities and *bridges* are selected from *gateways* using two marking and pruning schemes: static pre-pruning and dynamic pruning based on betweenness centrality [21].

Gently [42] is based on the Context-aware Adaptive Routing (CAR) protocol [44] and LABEL. LABEL will be the same as WAIT when a sender does not meet a member of the destination community (where the destination belongs to). In contrast to WAIT, Gently uses CAR-like routing to improve the efficiency when not meeting the destination community. Gently supports unicast as well as multicast and defines labels to identify communities of hosts. A recipient of message can be a single host or a group (community) of hosts. When the message carrier meets a node of destination community, Gently adopts a LABEL based strategy. Then in the destination community, CAR-like routing is used to deliver the message to the destination.

Zhou *et al.* [43] observed that while habitual mobility is useful in reducing the average communication latency, irregular deviation from habit can seriously affect worst-case performance. They proposed Diverse Routing (DR), a tunable protocol, to cope with the nodes deviating from their habitual activities. The main idea of DR is to statistically cluster the network into proximity-based social cluster and scatter at least one copy of a packet in a cluster such that even deviant nodes will be close to at least one of these packets. DR uses limited communication overhead to provide a stability and flexibility propagation scheme. Like Gently, DR supports unicast and multicast at the same time.

Community-based routing improves the forwarding efficiency depending on the higher meeting probabilities in the same community. Centrality and tie strength further provide evidence to choose the appropriate forwarder as well. However, distributed community detection and formation are still a challenge due to the dynamic topology of the network and difficulties in the information exchange and calculation. On the other hand, special nodes such as gateways have to be employed to accommodate inter-community communications. Hui and Crowcroft [38] introduced the concept of friendship community where two or more communities are very close or overlap and validated that the power law coefficient of friendship community is between intra-community and inter-community. Exploiting friendship among communities will be a potential way to improve the forwarding efficiency between communities.

*2) Community-Independent Routing*

Considering the difficulty of community detection and formation, many routing protocols have been presented without community support. The concept of ego network is

7exploited where only locally available information is considered [22][23]. Ego networks can be defined as networks consisting of a single actor (*ego*) together with the actors they are connected to (*alters*) and all the links among those *alters* [22]. These routing protocols are mainly based on utility function to calculate the satisfied nodes.

SimBet routing [23] is based on utility function which exploits betweenness centrality metrics and similarity to the destination node using ego network. In routing process, the encounter nodes firstly exchange the contact list to update the betweenness and similarity values. Then they exchange a summary list of destination values, calculate and compare the SimBet utility. If a node has a higher utility for a given destination, the destination is added to a request vector. Finally, they exchange the request list for further routing.

Further introducing the tie strength relationship with the destination, SimBetTS [22] was presented as an evolution of SimBet. Measuring tie strength in SimBetTS is an aggregation of a selection of indicators based on frequency, intimacy/closeness and recency. In the simulation experiments, betweenness utility, similarity utility, and tie strength utility have been examined respectively, as well as the SimBetTS protocol combining these three metrics. The results show that SimBetTS improves overall delivery performance while reducing the load on central nodes, which is better distributed across the network.

Some works embed context information of nodes into routing protocols. For instance, PROPHET [45] exploits the frequency of history contacts between users. MobySpace [46] and MV [47] exploit information about users' mobility patterns and places the users visit. The social relationships are important context information worth taking into consideration. HiBOp [48] automatically learns and represents context information, the users' behavior and their social relations, and exploits this knowledge to drive the forwarding process [6][76]. Nodes running HiBOp infer acquaintance between users through analyzing the similarity of their context information and behavior which include both present and historical information. Messages are forwarded through users closer to the destination. To improve the successful delivery rate, for each message more than one copy is injected into the network following a different route. To control the delivery cost, only the sender is allowed to inject multi-copies. In addition, HiBOp supports both unicast and group communications without requiring any particular customization.

Mtibaa *et al*. [49] exploited the information about social interactions of users from online social network platforms and applications such as Facebook, Orkut, or MySpace, etc. They proposed PeopleRank to rank the nodes in the social graph by using social relationships, inspired by the famous PageRank algorithm. Such social relationships can be based on explicit friendships (as defined in online social networks for example), on personal communications (for example extracted through the communication patterns available in cell phones), or even common interests. The node with a higher PeopleRank value will be more central in the social network and be a better forwarder. In addition, PeopleRank can be implemented in a centralized way or in a completely distributed fashion.

Aforementioned approaches take the social relationships among nodes into consideration. Several works leverage the regular human mobility pattern to predict the best forwarder. Some examples of predict-based routing protocols include PROPHET, MobySpace, MaxProp [50], etc. PROPHET and MaxProp are based on history contact data while MobySpace is based on the visit probability of locations in history. However, they all ignore the regularity of human mobility in spatial and temporal aspects which can take more accurate prediction for the future mobility.

In addition, the regularity of human mobility is considered in some works. PER [51] employs a time-homogeneous semi-markov process model to describe node mobility as transitions between possible locations. Location transition and sojourn time probability distributions are determined from nodes' mobility history. Nazir *et al*. [52] assumed that people follow similar mobility patterns daily (i.e. Monday to Friday) and proposed algorithms for social encounter based content delivery system with the time critical property.

The community-independent routings avoid the community detection, and leverage the context information, social properties (such as centrality, similarity, and tie strength) and the regular mobility pattern to predict the best forwarder. It is simple and easy to implement these routing protocols. One drawback of community-independent routing is that they are less sensitive to deviations. The prediction relies on historical mobility. If nodes deviate from their regular mobility patterns, the prediction will have large errors. For this reason, taking the deviation of mobility into consideration is a critical issue in the future.

Temporal and spatial factors are also of great significance. The history temporal/location information represent the people's movement pattern. The current temporal/location information indicate the instantaneous state which is related to the future state. Gao *et al*. took temporal factor into consideration in [53]. The transient node contact patterns are exploited. They define transient connectivity which indicates that some nodes may remain connected with each other during specific time periods and develop appropriate data forwarding metrics for more accurate prediction of node contact capability within the given time constraint. Existing works in the literature pay more attention to history temporal/location information, rather than current information. In addition, the combination of social properties and temporal/location information is worth investigating.

*B. Multicast Routing*

Multicast routing is a considerably fresh and challenging problem in frequently disrupted and partitioned networks. Multicast routing allows information to be delivered to multiple destinations. Usually, the multiple destinations are identified as a group. Group membership of a particular multicast group may change with time as nodes join and leave the group [54]. Patra *et al*. [55] classified multicast routing strategies in DTNs on the basis of their basic working mechanisms. The categories include flooding multicast routing, tree-based multicast routing, probability-based multicast routing, and intelligent multicast routing. None of them consider social properties.

Inherent social relationships group people into communities, which can be leveraged for the purpose of multicast routing. Some works aforementioned in unicast routing support multicast such as community-based routing. In community-based routing protocols, if the recipients of the multicast are in the same community, it is feasible to support multicast. The process is that the message is forwarded to any member of the community and then replicated in an epidemic manner among the members of the community. Taking HiBOp as an instance, it supports context-aware multicasting using the destinations' identity table.

Gao *et al.* [54] are among the first to study multicast from the social network perspective. They proposed a set of multicast routing methods with single and multiple data items based on community and centrality. In single-data multicast, they introduced a new metric called cumulative contact probability (CCP) to indicate the average probability that a node meets a random node within a time constraint. Based on CCP of nodes, the minimum number of relays is selected (by solving a unified knapsack problem) to satisfy the requirement on delivery ratio within certain time constraint. In multi-data multicast, a node maintains its destination-awareness about other nodes to select relays among its neighbors. If the destinations are in other communities, data forwarding is conducted through gateway nodes to connect multiple communities. The multi-data multicast is also modeled as a knapsack problem ensuring the required delivery ratios.

In multicast, the selection of relays is usually based on the relays' cumulative probabilities in order to forward single or multiple data to multiple destinations [54]. However, unicast generally relies on the future meeting probability to decide the relay. Compared to unicast, multicast protocols face more complicated context and are more challenging to design. However, multicast has a wide spectrum of application scenarios which require sharing data among people in e.g. a meeting, conference or battlefield scenario.

### C. Congestion

One critical issue of routing protocols is the trade-off between improving delivery ratio and reducing delivery delay and overhead. The socially-aware routing protocols mentioned above contribute to this target by selecting appropriate forwarders which have more opportunities to encounter the destination from the perspective of social interactions. However, the problems of overhead and congestion remain to be addressed. In the context of networks, congestion generally refers to a network state where the relevant node is overloaded with too many messages. A result of congestion is that network QoS may be deteriorated. The cause of overhead and congestion may come from e.g. the redundant message duplications and over-use of the central nodes. The physical buffer of any mobile node is limited. When there are excessive messages to be exchanged, congestion may happen [56].

Redundant message duplications might cause high overhead and congestion of the network. In the worst case, 94% of duplicate packets reach the destination, which will induce huge overhead on bandwidth, energy and memory consumption [57]. Many protocols are trying to achieve an efficient trade-off by controlling duplication.

Kawarabayashi *et al.* [57] proposed a strategy to lower message duplication under a given delay or delivery probability. Based on predictable working day model, the authors formulated the problem of message duplication into a spanning tree problem which is further optimized in delivery time, duplication and storage space of messages. In [58], the delegation forwarding algorithm forwards a message only if it encounters another node whose quality metric is greater than any other nodes seen by the message so far. The cost of delegation forwarding algorithm is proportional to the square root of population size, which is more optimal in contrast to naive forwarding with linear cost.

Most of the routing protocols lead the routing to direct most of the traffic through a small subset of good nodes (e.g. central node). For instance, in the SimBet algorithm, the top 10% of nodes carry out 54% of all the forwards and 85% of all the handover [59]. This unfair load distribution causes local storage congestion, further increasing the discarding rate and decreasing the delivery rate.

To deal with congestion problem, current methods mainly focus on two strategies. One is to mitigate the role of the central nodes. Fair routing [59] classifies the nodes into different levels to limit the exchange of messages. It exploits the social process of perceived interaction strength based on the social relation between nodes in different time scales. Then it forwards the message by the stronger social relation and uses assortative-based queue control to limit the exchange of messages to those users with similar "social status". In other words, nodes will only accept forwarding requests from those nodes of equal or higher status. In addition, Yuan *et al.* [62] introduced strangers to participate in routing and exploited the optimized number of strangers in order to achieve a better performance of overhead/packet delivery ratio in pure darkness scenario. The other strategy is based on detection of the congestion. Nile [55] keeps the link loads in check to push replicas only to those promising paths that may sustain more loads. Radenkovic and Grundy [60] introduced congestion driven part in forwarding protocol to avoid nodes that have lower availability and higher "congestion rates". In reality, the detection of congestion and the prediction of idle paths are difficult to address due to changing network topology. Recently, some researchers tried to make use of strange nodes and nodes with weak ties, which might be a promising approach to cope with congestion.

Table I summarizes representative socially-aware routing and forwarding protocols in the literature. The socially-aware routing and forwarding protocols using community can support both unicast and multicast, which have more flexibility and applicability. A lot of experiments and simulations have been carried out in the protocols listed in the table which are usually based on real trace data. The results of experiments and simulations show that the

4TABLE I
SUMMARY OF SOCIALLY-AWARE ROUTING AND FORWARDING PROTOCOLS

| Solution | Social Characteristics | Unicast or Multicast | Trace-Based Mobility | Protocols for Comparison |
|---|---|---|---|---|
| LABEL [33] | community | unicast, multicast | InfoCom06 | MCP, WAIT, Control |
| BUBBLE RAP [19] | community, degree centrality | unicast, multicast | HongKong, Cambridge, InfoCom06, MIT Reality Mining | MCP, LABEL, FLOOD, WAIT, PROPHE |
| LocalCom [16] | community, tie strength, betweenness centrality | unicast, multicast | Haggle, MIT Reality Mining | PROPHET, BubbleRap |
| Friendship-based [36] | community, friendship, similarity, tie strength | unicast, multicast | MIT Reality Mining | PROPHET, SimBet |
| Gently [37] | community | unicast, multicast | Implemented in Haggle | |
| DR [38] | community | unicast, multicast | Duk trace, WTD trace | SimBet, Epidemic |
| SimBet [18] | betweenness centrality, similarity | unicast | MIT Reality Mining | PROPHET, Epidemic |
| SimBetTs [17] | betweenness centrality, similarity, tie strength | unicast | Intel, Cambridge, InfoCom06 | PROPHET, Epidemic |
| HiBOp [43] | context information, social relationship, similarity | unicast, multicast | Community based mobility model | PROPHET, Epidemic |
| PeopleRank [44] | friendship, similarity | unicast | MobiClique, SecondLife, InfoCom06, Hope | |
| Social-based [49] | community, centrality | multicast | MIT Reality Mining, InfoCom06 | Epidemic, BubbleRap, PROPHET, SimBet |
| Fair Routing [52] | community, centrality | unicast | MIT Reality Mining | Epidemic, PROPHET, SimBet |

socially-aware routing and forwarding protocols outperform socially-oblivious protocols, as a general rule.

## V. ROUTING AND FORWARDING WITH SELFISHNESS

All the approaches mentioned above assume all nodes in the network are cooperative and altruistic. Therefore, all nodes are willing to help forward messages for other nodes. But in reality, many nodes exhibit non-cooperative behaviors, such as selfish nodes or even malicious nodes, in order to e.g. conserve limited resources (like power and buffer) and increase their own benefits.

A selfish node always acts for its own interest, meaning selfish nodes may not be willing to provide services for others in order to conserve their limited buffer or power resources. Whereas, a malicious node acts maliciously with the intention to disrupt the main functionality of the networks, so it might possibly drop packets, jam the wireless channel, and even forge false packets [63].

Many efforts have been made in the literature to evaluate the effects of cooperation and/or selfishness in opportunistic networks from different aspects [64]-[69]. It has been proved that node collaboration (even limited collaboration) can dramatically improve performance compared to non-cooperation scenarios and different non-cooperation behaviors may have opposite impacts on different routing algorithms.

Social selfishness is first introduced in [70]. In social perspective, a selfish user is usually willing to help others with whom he/she has social relationships (e.g., friends, coworkers, roommates) or in same community. He/she will provide better service to those with stronger social ties than those with weaker ties, especially when there are resource constraints. The social related selfishness is defined as social selfishness. In contrast, individual selfishness refers to the nodes having the same selfishness level to other nodes. Social selfishness conveys the social tie between the nodes and can be used to select trusted relay nodes. When a node has no social ties to the outside world, e.g. a node is out of its own community, it becomes individual selfishness. It would be better to treat nodes' selfishness behaviors in different ways with respect to different scenarios.

To deal with selfish nodes, incentive mechanisms are necessary to stimulate nodes' cooperation, check misbehaviors and punish selfish nodes. Due to frequent network partitions, resource constraints and longer delay, the incentive mechanisms developed in wireless networks like MANETs, may not work effectively in intermittently (dis)connected scenarios [68]. This is a challenging problem that has attracted a lot of attention from many researchers.

Most recent works addressing node selfishness fall into one of the following four categories: reputation-based, credit-based, Tit-For-Tat (TFT), and trust-based. Generally speaking, the first three categories are traditional incentive mechanisms that focus on individual selfishness, while trust-based strategies take social selfishness into consideration. One could argue that trust-based solutions do not belong to incentive mechanisms because they do not take any action to stimulate cooperation or check the selfish node. In the context of socially-aware networks, however, social selfishness becomes an important issue to be addressed. Consequently, we cover trust-based strategies here. Different categories of solutions have different principles to stimulate cooperation. At the same time, they need to cope with various attacks coming from malicious nodes.

## A. Reputation-Based Incentive Mechanisms

In general, reputation-based incentive mechanisms identify misbehaving nodes and exclude them from the network. Nodes can build up their good reputation scores by forwarding packets for others, and thus will be rewarded with higher priorities when transferring their own packets. The corresponding reputation score decreases when a node misbehaves. The node with bad reputation is detected and excluded when its reputation becomes below a threshold [74].

In reputation systems, sybil attack and whitewashing attack are common attacks to be handled. A sybil attack [71] indicates that a malicious node attempts to create multiple identities in order to help itself gain good reputation, abandon bad reputation or evade responsibility of its actions. In a whitewashing attack [72], a node repeatedly leaves and rejoins the system using newly created identities to avoid suffering from bad reputation. In classical networks, trust is established by a certificate authority (CA) through a public key infrastructure (PKI) [73]. But it is not easy to implement in non-infrastructure and intermittent connection scenarios.

RADON [74] is a reputation-assisted data forwarding protocol which comprehensively evaluates an encounter's forwarding ability by integrating the reputation of forwarding data with the possibility of meeting a destination. In RADON, a special message called Positive Feedback Message (PFM) is used to help the Watchdog monitor a node's forwarding behavior in the reputation management system. RADON prevents a malicious node from deliberately dropping and arbitrarily forwarding data, which dramatically improves the network performance in a malign environment.

Give2Get Epidemic Forwarding and Give2Get Delegation Forwarding are the first protocols for packet forwarding in a social setting to tolerant selfish behaviors, which are based on Epidemic forwarding and Delegation forwarding respectively [74]. Give2Get consists of two phases: the relay phase and the test phase. In the relay phase, the sender (S), which generates the message, tries relaying the message to (at least) the first two nodes it meets by negotiating a cryptographic session key. Assume node B relays the message, node B continues the relay phase to find two other nodes and relay the message to them. By doing this, it can collect two proofs of relay for itself. When meeting node S again, during the test phase, if node B is not able either to show the two proofs or to prove to have still the message in its memory, then node S can broadcast a proof of misbehavior (PoM) to the whole network, which will remove node B if it cannot prove that node S is wrong. In addition, both protocols are Nash equilibriums, i.e. no individual has an interest to deviate.

MobiID [76] is a "user-centric" and dynamic reputation based incentive scheme to stimulate cooperation in bundle forwarding. Conventional reputation based schemes always rely on neighboring nodes to monitor the traffic and keep tracks of each other's reputation. In contrast, MobiID allows each node to maintain, update and show its reputation tickets by itself on demand. This is called self-check. It also defines a social metric which considers the forwarding willingness from forwarding history and identifies the social community based on this metric. Through sharing reputation information in the community and forming consensus views towards targets, MobiID implies community check. MobiID exploits Offline System Manager (OSM) which is responsible for key distribution. In addition, MobiID addresses attacks such as modifying the forwarding history to over-claim a high reputation (so as to attract bundles) and isolate the node from the target user.

IRONMAN [77] is another incentive mechanism using social information. However it uses pre-existing social information to detect and punish selfish nodes, thus incentivizing them to participate in the network. The social information can be obtained through interview, or from an online social network (e.g., Facebook friend lists). With the proliferation of online social network applications such as Facebook, Orkut, or MySpace, the social interaction information of users becomes available. The social information is increasingly reliable and can be utilized for many purposes such as predicting future encounters in opportunistic networks. The integration of online social network and mobile networks is a prominent problem in future research work.

SRed [78] is a localized, link-state-based and multi-path routing protocol which also mitigates a number of known routing layer attacks such as black hole, denial of service (DoS) and wormhole. SRed uses reputation-based routing generation mode in benign environments, otherwise the probabilistic routing generation mode will be activated. The dynamic window mechanism is used to switch between the two modes in order to achieve a good trade-off between efficiency and security.

The above protocols avoid central management mechanisms to control the reputation estimation in order to adapt to mobile environments. RADON, Give2Get and SRed utilize the successful forwarding process to verify misbehaving nodes. Li and Cao [79] presented a similar scheme to migrate routing misbehavior through detecting packet dropping.

All these protocols can achieve high delivery efficiency, but they benefit from different aspects. RADON, Give2Get, SRed and Li and Cao's scheme are easy to operate, while MobiID and IROMAN are of high reliability thanks to the exploring of social community and group strength.

## B. Credit-Based Incentive Mechanisms

Credit-based incentive schemes introduce some forms of virtual currency or credit to regulate the packet-forwarding relationships among different nodes [80]. Nodes earn credits by forwarding packets. As a return, these credits can then be used to obtain forwarding service from other nodes in the system. That is, it will take some credits for the source node to send each packet. If a packet is delivered successfully, the corresponding credits will be distributed to the intermediate nodes participating in packet relay.

There are two significant difficulties in credit-based incentive schemes. One is the management of nodes and credits distribution. Sometimes, the realization of payments requires an out-of-band trusted third party, which is unlikely to be available in DTN. The other is cheat attacks from selfish nodes. Due to the selfish nature, mobile nodes will try to cheat the system to maximize their welfare through injecting or deleting some relay nodes to achieve more credits.

In SMART [80] and Pi [81], the systems employ an Offline Security Manager (OSM), which is responsible for key distribution, and a Virtual Bank (VB), which is responsible for credit management. SMART is a secure multilayer credit-based incentive scheme for DTN which can be compatible with diverse data forwarding algorithms. It is based on the notion of a layered coin that consists of one base layer and several endorsed layers. The base layer is generated by the source while the endorsed layers are generated by forwarding nodes. This layered coin mechanism makes SMART withstand a wide range of cheating actions (or attacks) such as layer injection attack, nodular tontine attack and submission refusal attack. Pi is a variant of SMART which combines reputation- and credit-based schemes. In Pi, if and only if the messages arrive at the destination, the forwarding nodes can get credits from the source node. To achieve fairness, the forwarding node still can get good reputation from a trusted authority (TA) for the failure forwarding.

MobiCent [73] also makes use of a Trusted Third Party (TTP) to store key information for nodes and provides verification and payment services. It uses incentive-compatible payment mechanisms to cater to client that wants to minimize either payment or data delivery delay and handle the edge insertion and edge deletion attacks. In this scheme, nodes are paid for forwarding packets and the destination makes the payment decision. No node will get incentive to tamper with the path it reports to the destination.

### C. TFT Incentive Mechanisms

TFT is based on the basic principle that "I'll do for you as much as you did for me" [84]. BitTorrent [82] is one of the most popular P2P (peer-to-peer) systems using direct TFT reciprocity strategy. In BitTorrent, a user's download rate is proportional to its upload rate. The TFT incentive scheme often suffers from bootstrapping problems and injecting of fake messages.

Incentive-Aware Routing [83] is a pair-wise TFT incentive routing protocol which combines generosity and contrition. The generosity addresses the bootstrapping and absorbs transient asymmetries, while contrition prevents mistakes from causing endless retaliation.

Buttyán et al. [84] proposed a mechanism to discourage selfish behaviors based on the principle of barter: a user who trades in messages can download a limited volume of messages from another user if he/she can give the same volume of messages in return. In this scheme, digital signature and reputation mechanism can be used to prevent the injection of fake messages.

RELICS [85] is a mechanism to combat selfishness in energy-constrained DTNs. The incentive mechanism used in RELICS is called reciprocity of service. Every node is given an explicit rank based on its transit behavior (i.e., forwarding messages originated from others). The rank of a node is accumulated when the node participates in relaying messages, whereas the rank is decreased when the node sends message. Based on the rank of the source node, messages originated from highly-ranked nodes are given priority over lowly-ranked ones. Furthermore, RELICS, for the first time, takes energy into consideration for the incentive mechanism in DTNs. In the scheme, it considers energy to be the core rationale behind selfishness, and each node is allowed to set a delivery threshold and to adapt its energy depletion rate based on its rank such that the rate is merely enough to achieve the desired delivery ratio.

Existing credit-based and TFT incentive mechanisms do not take social properties into consideration directly. However, as traditional incentive mechanisms, they can be applied to social-based routing protocols. Furthermore, credit-based and TFT incentive mechanisms with social awareness will be interesting research topics.

### D. Trust-Based Strategies

Social selfishness is closely related to not only the willingness intention for forwarding but also the trusted relationship between nodes. Furthermore, more relationship trust implies stronger social tie between nodes, which can be used in effective relay node selections during the course of forwarding. Trust-based strategies often establish trusted relationship to complete trusted routing by coping with social selfishness.

Considering social selfishness and following the philosophy of *design for users*, SSAR [70] allows user selfishness and improves performance by considering user willingness, resource constraints, and contact opportunity when selecting relays. User willingness values can be configured via user interface in the mobile device.

Chen et al. [63] integrated social trust and QoS trust into a composite trust metric for determining the best message carrier among new encounters for message forwarding. They consider healthiness and cooperativeness for social trust to account for a node's trustworthiness for message delivery, and connectivity and energy for QoS trust to account for a node's QoS capability to quickly deliver the message to the destination node. Trust-threshold based routing (TTBR) was proposed which designs trust thresholds for determining the trustworthiness of node acting as a recommender or as the next message carrier. TTBR is distributed and does not require a complicated credit management system.

Considering the social network structure and its dynamics, Trifunovic et al. [86] proposed two approaches for social trust establishment that are robust to sybil attacks: explicit social trust and implicit social trust. Explicit social trust is based on consciously established friend ties by building a robust tree-like graph of paired users. Implicit social trust leverages mobility properties using complex network tools and builds another graph up to two-hops based on the familiarity of surrounding peers and the similarity to reinforce trust in a user.

Mtibaa and Harras [87] leveraged social information and proposed six trust based filters to establish trustworthy communications over mobile opportunistic networks. The six filters couple three socially-aware estimators of trust including common interests, common friends, and the distance in the social graph, with two major techniques of trust establishment including Relay-to-Relay and Source-to-Relay. It has been shown that the trust filters yield a fair trade-off between trust and success rate.

Trust-based strategies provide more effective and secure forwarding solutions. But the establishment of trust may face the difficulty that centralized mechanisms cannot be easily deployed in infrastructure-less networks such as opportunistic networks. It is also challenging to efficiently and effectively propagate the social information.

## VI. Socially-Aware Data Dissemination

Large volume of data is being generated every day. Many content-based services or information are on the fly in the mobile networks. Meanwhile, with the emerging User-Generated Content (UGC) service, the users are not only the consumers but also the producers of the content. The content-based services wish to push the data to their describers, while people wish to produce and share the content with their friends. Consequently, there is an increasingly crucial demand to data dissemination in multi-point asynchronous manner in practice. With multi-point asynchronous communication, the destination of communication can be a group of nodes, and communications between nodes use asynchronous way. From the social perspective, people always group into communities and their behaviors are regular. These social properties are beneficial for improving the efficiency of data dissemination.

The Publish/Subscribe (Pub/Sub for short) paradigm recently emerges as a promising solution to data dissemination thanks to the decoupling characteristic. It has attracted a lot of attention from many researches. The decoupling characteristics give the advantage of removal of static dependencies in a distributed environment, which is beneficial in supporting highly dynamic and decentralized systems [25].

In Pub/Sub paradigm, three roles are usually deployed: publisher, subscriber and broker. The publisher is the information producer which issues newly-detected events without having to specify the receiver. The subscriber is the information consumer which expresses their interest in certain events without knowledge of what the publisher might be. The broker is the interface between the publisher and subscriber which provides routing, event matching, and filtering services, etc. In the context of data dissemination, the nodes providing the contents are publishers. The nodes that are interested in the contents are the subscribers. The relay nodes are the brokers. As a consequence, mobile nodes can be publisher, subscriber or broker alternatively [25].

Recent works about socially-aware data dissemination fall into one of two categories depending on different point of view. The first category is based on solicitation and cache, from the viewpoint of nodes. For a node in this context, besides its interested content, it can cache the uninterested content for others to improve the dissemination efficiency. However, the buffer size of any node is limited. It is impossible to store all encountering contents. Frequent buffer replacement will induce considerable consumption of power and decrease the efficiency of the network. Thus, soliciting and caching appropriate set of contents for future distribution according to local and/or global environments is an effective way to improve the efficiency of data dissemination and to reduce energy consumption. The second category is based on forwarding, from a perspective of the content. People sharing common interest and/or activities can build communities which facilitate communications and information sharing between them. In this regard, how to find the proper forwarders to carry them to the destination community as quickly as possible is the core of successful data dissemination.

In this section we will give an overview of solicitation and cache based approaches, and forwarding based approaches respectively. And most algorithms in these two research directions employ similar mechanism with PodNet [88]. Therefore, we will give a simple introduction to the PodNet firstly. In PodNet, the improvement approach for content distribution focuses on solicitation and five solicitation strategies are proposed. Accordingly, we classify PodNet into the first category.

### A. Solicitation and Cache Based Approaches

The PodNet project is a Pub/Sub paradigm application for data dissemination in wireless ad hoc networks. In PodNet, contents are organized into feed channels. Users subscribe to channels they are interested in. Users are associated in pair-wise way when they come into the transmission range of each other. The behavior of the scheme can be described as a receiver-driven broadcasting paradigm. The intermediate nodes are allowed to solicit and cache the unsubscribed contents (i.e. uninterested contents) according to their available buffers so that it can better serve its potential encounters in the future [89].

With PodNet, the cache structure of a node is separated into two parts: private cache and public cache. The private cache is used for storing subscribed channels and the public cache is used for public channels that are solely for redistribution. It is assumed that the capacity of private cache is large enough to store the interested contents while the public cache is limited. Upon pair-wise contacts, nodes first exchange the subscribed content which is matched using Bloom Filter. Then nodes use the remaining connection time to update and download contents for public cache by using certain solicitation strategy. Five solicitation strategies are presented, including Most Solicited, Least Solicited, Uniform, Inverse Proportional, and No Caching. These strategies are based on popularity of content which can be obtained and maintained locally by exploiting the information from the requests of other peers except for the No Caching strategy. For example, Most Solicited caches the most popular content while Least Solicited caches the least popular content.

Due to buffer size limits, there are two key issues with the Pub/Sub paradigm, including soliciting proper unsubscribed content to cache and replicating proper unsubscribed content when the buffer is full.

Ma et al. [90] proposed to make a soliciting and caching decision by jointly considering the history encounter information, the content preferences of the subscribers and the popularity of the contents. Using these information, each node evaluates all contents waiting for caching and solicits those unsubscribed contents it prefers to cache for future distribution.

Chuah et al. [91] proposed a data model that categorizes data into sets. Each category has a certain set of associated keywords. Users' interests are described by using the same keyword space as the data model. When two nodes encounter, each node exchanges the meta-data descriptions of stored data and counts how many nodes obtain data from itself in an observation window. If the data is of interest, the node will request for it. Otherwise, a node will request for data of no interest only with a probability which is dynamically adjusted based on the average rewards achieved from storing data of this category in the previous observation window.

In the above-mentioned approaches, every node makes independent dissemination decisions. Ma et al. [92] presented a cooperative cache-based data dissemination framework (CCCDF) to carry out the cooperative soliciting and caching strategies for encountering nodes. Considering two encountering nodes as a group, CCCDF produces cooperative dissemination decisions based on all relevant information such as nodes' mobility characteristics and content subscriptions. Through updating and exchanging average meeting rates, two nodes solicit the subscribed and unsubscribed content download sequence as well as drop sequence in case the cache is full. Based on the CCCDF, Ma et al. proposed two cooperative content dissemination strategies for different motivations [92]: CCCDF (Optimal) is an optimal strategy to maximize the overall content delivery performance while CCCDF (Max-Min) is a cooperative strategy to share the limited network resources among the contents in a Max-Min fairness manner no matter how popular the content is.

Considering that mobile nodes often move to places where they can establish communications, Jaho and Stavrakakis [87] integrated interest social group with locality-induced social group to enhance data dissemination. They compared two storage strategies: selfish and cooperative. When soliciting contents, the cooperative strategy takes into account the interests of the nodes they will (most) likely encounter in the future besides its own interests, whereas the selfish strategy only caches contents of its own interest. It has been shown that the cooperative strategy outperforms the other one.

ContentPlace [94]-[96] employs the same community detection mechanisms as [25], but does not need any overlay infrastructure. ContentPlace is not limited by the assumption that the members in a community have common interest. In contrast, it assumes that different communities have different interests and one community have different interests either. ContentPlace considers the relationship of one user with several communities, that is to say, users belong to several different social communities. It assumes that users automatically learn the time spent on the communities, the interested data types and the data spread in communities. These kinds of information are used to evaluate the utility of each encountered data. The core of ContentPlace is to select a set of data objects to cache in order to maximize the local utility of its buffer by solving the multi-constrained 0-1 knapsack problem. In addition, ContentPlace also considers five policies to evaluate the social weight, i.e. Most Frequently Visited (MFV), Most Likely Next (MLN), Future (F), Present (P), and Uniform Social (US). The simulations provide best results in the Most Likely Next and Future policies.

*B. Forwarding Based Approaches*

The key factor of forwarding based approaches is how to select appropriate nodes in order to complete the data dissemination as quickly as possible. Human beings' social properties are beneficial to predicting the information of destination and meeting probability. Community and centrality are two popular properties in socially-aware data dissemination. In the first place, people with common interests always come into a community and there are some "central" people in a community. The communities are the sole destination of relevant contents and can help the content find their destination easily. Central nodes can help the content travel quickly in community. Additionally, people always follow regular mobility patterns. They always access several locations frequently. The mobility regularity can help to select appropriate relay nodes to minimize the intermediate nodes and improve the efficiency of data dissemination.

In [25], a socio-aware overlay is built for Pub/Sub communications by using the community and centrality concepts. It detects community dynamically in which community members are well connected, implying that socially they share the same interests with high probability. Nodes with high closeness centrality in communities are selected as brokers to construct an overlay to facilitate multi-point Pub/Sub event dissemination. Since nodes with high closeness centrality have the best visibility to other nodes in the community, it will be relatively reliable to delivery contents to any (other) member of the community through these nodes. However, it is hard to maintain the overlay network because nodes with highest centrality values may change over time (due to e.g. mobility).

SocialCast [97] is an interest-based routing framework for Pub/Sub paradigm and it is based on utility function to select best carriers. The approach assumes that socially-bound hosts are likely to be co-located regularly. It exploits prediction based on metrics of social interaction (probability of a user to be co-located with another one sharing the same interests and change degree of connectivity) to identify the best information carriers, which implies the dissemination of messages through matching the subscriber's interests. SocialCast asks a publisher to insert *r* copies. When a better carrier is encountered, only one copy is removed from the local buffer and sent to the new carrier. Therefore, at any time, the network contains at most *r* copies of the message. This approach to message distribution is also explored by other solutions such as Spry and Wait [98]. SocialCast works well when a community is interested in the same type of contents, but it is not clear how it works in more general settings.

Li and Wu [26] proposed a Mobile cOmmunity-based Pub/Sub (MOPS) scheme which utilizes the long-term social network properties to facilitate content-based services. MOPS defines the closeness metric based on nodes' temporal and spatial encounter information to depict the neighboring relationship between nodes. The closeness-based local community is defined as a clique of nodes where any neighboring relationship is stronger than an adjustable threshold. It considers not only the direct but also the indirect neighboring relationship. MOPS combines push and pull strategies and determines the boundary between communities as the push-pull boundary. Within the community, nodes broadcast interests and publishers send contents directly to the nodes. Brokers are then deployed as the interfaces to match the interests and bridge communities.

Habit [27] leverages information about nodes' regularity of movement and their social network (or network of interest) to construct regularity graph and interest graph respectively. The regularity graph is responsible for describing the familiar stranger relationship between nodes with temporal and spatial relevant. If a node encounters anther node frequently enough, it becomes a familiar stranger in a regular time period. The interest graph is responsible for describing the relationship of interested content transmission between nodes. Using these two graphs, Habit enables relevant content to reach interested nodes while minimizing the computation and communication load on uninterested intermediate nodes, while still achieving a high delivery rate.

Fan *et al.* [99] addressed data dissemination among several communities and proposed to make decision on the routing trajectory based on semi-Markov analytical model. They defined *geo-community* based on the geographic regularity of human mobility and *geo-centrality* as the super user among communities. They formulated the super user route design as a combinational optimization problem of Convex Optimization and Traveling Salesman Problem to achieve the goal of minimizing the total duration and guaranteeing the required data dissemination probability.

Forwarding based approaches exploit social information in the working environment and take into consideration social relationships between users to select the data object to exchange. The networks are generally divided into several communities that provide more global information. However, due to the dynamic topology of the network, more resources will be consumed to detect and maintain the communities. In addition, there is no consideration of the multi-interest scenario. One challenge is that we must find a way to achieve a fairly distributed delivery rate.

The main difference between solicitation and cache based approaches and forwarding based approaches lies in what they concern about. Solicitation and cache based approaches emphasize more on messages' properties. They usually consider messages' popularity and dissemination rate. Thus they are suitable to content-aware environments. For instance, the community is detected according to individuals' interest. While forwarding based approaches emphasize more on individual's properties. They pay more attention to the individual's status in order to disseminate faster. Therefore, they have been more widely adopted.

In addition, the cache freshness problem is attracting researchers' attention. Some messages such as news are time-sensitive. As a result, to achieve cache freshness, the date should be the newest in cache. However, previous researches usually maintain the freshness of cached data by refreshing periodically. Recently, Cao *et al.* [100], for the first time, proposed a scheme to efficiently maintain the cache freshness by organizing the caching nodes as a tree structure during data access. Each node in the tree is responsible for refreshing the data cached at its children in a distributed and hierarchical manner.

## VII. OPEN ISSUES

In previous sections, we have reviewed the state of the art in the emerging field of SAN. Accordingly, we try to cover the majority of categories of related research efforts, though the content is not projected to be exhaustive. From the analysis and comparison of these methods, we can conclude that the social properties are the potential improvement strength to mobile network design. There are large spaces to improve the efficiency of protocols and algorithms by exploring comprehensive use of social properties. In addition, although security and mobility model from social aspects are not involved in this paper, they are also hot research topics in SAN and many research works have been conducted, e.g. [63], [101]-[112]. Besides, quite a number of research projects and initiatives closely related to SAN have been launched in recent years. A list of these research projects is provided in Table II. Despite the considerable amount of ongoing research, the SAN research community is very young. Many challenges remain yet to be addressed. On the other hand, there are a lot of opportunities to improve the efficiency and effectiveness of SAN to accommodate new applications and services. In this section, several open issues will be outlined.

### A. Mobile Social Sensing and Learning

In the context of SAN, protocol and algorithm design takes into account relevant social properties as main basis. It is necessary to sense and learn the context information in order to obtain the social properties. In present works, the social properties mainly come from the analysis of the experiment data and most existing studies are based on analytical or simulation models. Then comes the challenge that how to bring SAN into real applications with mobile social sensing and learning.

Basically, it might be easy to obtain raw data of the context information such as time, location, action, etc. by using mobile devices equipped with appropriate sensors. In this aspect, most of the projects listed in Table II have provided large amounts of datasets of inevitable value. While, since majority of the existing data is about information like time, location, which is not enough for further analysis on social behavior, novel mobile social sensing technologies need to be tested, especially for collection of users' words, actions, etc.

TABLE II
SUMMARY OF SOCIALLY-AWARE NETWORKING RELATED PROJECTS

| Project | Organization(s) | Research Area(s) | URL |
|---|---|---|---|
| SocialNets | FET | Social networks for the delivery and acquisition of content covering online opportunistic wireless network | http://www.social-nets.eu/ |
| PeerSoN | KTH, EPFL, NTU Singapore, etc | P2P infrastructure supporting features of online social networks. | http://www.peerson.net/ |
| MetroSense | Dartmouth College | Applications, classification techniques, and sensing paradigms for mobile phones capable of societal-scale sensing | http://metrosense.cs.dartmouth.edu/ |
| A Million People | | Mobile network, epidemiology, urban planning, and social science. | http://www.amillionpeople.net/ |
| PodNet | CSG ETH Zürich | Content distribution protocol and system on mobile devices. | http://podnet.ee.ethz.ch/ |
| GroupMedia | MIT | Perceptual socially-aware applications for cell phones and PDAs. | http://groupmedia.media.mit.edu/ |
| Haggle | FU FET | Situated and autonomic communications | http://www.haggleproject.org/ |

After being collected, the useful information needs to be extracted from the sensor data for real usage, as a result of the fact that, different applications need different social properties and different social properties require different data at different scales. For example, the human mobility pattern is sensitive to the time and location at individual scale, while the community and centrality are sensitive to the interest or interaction at group scale. This makes the mobile social learning a vital issue. In other words, an SAN application should be able to analyze the collected data or classify them according to the social features they reflect. In some occasions, the application should also be able to predict the trend or future pattern to afford users a smarter life, for example, remind a meeting, help schedule, or make an appointment. To this end, the mobile devices need to support large storage capacity and computation capacity. In this regard, mobile cloud computing is a suitable solution to computing and mining big data for very large numbers of users [101], with the ability of providing available computing and storage resources.

Based on the above statements, both mobile social sensing and learning need further research on improving their own effectiveness and efficiency considering the points of application and social properties.

*B. Privacy*

Mobile devices, which are used to monitor people's context information, record their preferences and behaviors, and hence possess the users' private information. Most of the protocols and algorithms for SAN require users to share their personal information such as physical location [114], preference, and social relation. Therefore, the privacy issue for mobile users in SAN becomes crucial. A generalized privacy policy is hard to implement because the granularity of privacy concern may differ from user to user, even the same user has different sensitivities to the same information in different applications [115].

The solutions to privacy presented by existing works are usually encryption or access control to private information. Encryption only exposes the sensitive information to the identified users, which is not suitable to SAN due to the general requirement of central server. Information access control allows users to disclose their private information to "close" users. For instance, most mobile social networking services allow a user to share his/her location with his/her friends, even friends of friends (FoF). In the context of SAN, access control to private information becomes difficult due to various relationships involved. There is a tradeoff between obtaining accurate social properties for socially-aware networking protocol design and protecting privacy.

*C. Node Selfishness*

Dealing with selfish nodes is an important but challenging issue in SAN. As mentioned before, node selfishness in this context can be classified into social selfishness and individual selfishness. Many research efforts on incentive mechanisms are exploited to stimulate the individual selfish nodes to cooperate [116]. However, the social selfishness is usually used to help select trusted nodes in routing protocols and they are not exploited sufficiently, particularly in incentive mechanisms. It is necessary to detect and deal with the individually selfish nodes and socially selfish nodes differently according to the context information, particularly social information. Addressing the problem of selfish nodes effectively will be beneficial to not only the design of routing protocols and data dissemination algorithms, but also handling of privacy and security problems. As a consequence, node selfishness is worth investigation.

*D. Scalability*

Scalability is an open problem that is seldom considered in previous works [117]. Indeed it should be taken into account on each level of SAN's architecture. Here we take the routing and data dissemination, and application levels as examples.

On the level of routing and data dissemination, most of the existing protocols and algorithms require nodes to store their encounter information with others or context information to estimate the appropriate relays. However, mobile devices are limited by the wireless spectrum and onboard resources, especially the low buffer storage and energy. This causes the problem of scalability and makes the routing protocols or data dissemination algorithms badly compatible with the increasing node density of the network in many aspects: (1) If new nodes join the network or some nodes are powered off, a new route between two nodes needs to be detected, and this costs energy and time; (2) When the network is large, finding an effective route will be difficult, especially in situations where the topology is dynamic; (3) Larger memory space is needed to store the increased encounter information; (4) Sometimes, there are multiple feasible routes to disseminate the information; in this case, optimal route selection and route management become challenging problems. All these points require mobile devices to have good capacity in terms of buffer, computation and energy efficiency.

On the level of application (or system), scalability is also an important issue. For a robust system, the ability of being able to be extended with new features easily is very necessary when facing with changing requirements. When a new device joins the system, it is necessary to integrate the device into the system quickly, which is also an essential aspect of scalability. Besides these scenarios, it is possible to identify many requirements in terms of scalability in the context of SAN.

## VIII. CONCLUSION

Socially-aware networking is going to be a new hotspot of research on network science and engineering. The close connection between ubiquitous mobile devices and the users' social relationships attracts researchers to explore the potential of introducing social properties into network design. We believe that SAN will benefit the engineering of next generation networks as a promising paradigm.

In this paper, we have discussed basic concepts behind this new terminology and presented a first survey of state of the art. Through examining existing research results, it can be seen that social properties are indeed a powerful source not only for the design of outstanding networks, covering the areas of routing and forwarding protocols, but also for tackling the problems related to selfishness behavior in routing and forwarding situations. Additionally, we described the major approaches on socially aware data dissemination to make full use of the social properties. However, a multitude of challenges remain to be addressed before the full potential of SAN can be realized in practice. We have examined some of these open issues in this paper.

We hope this survey will provide a better understanding of the literature of SAN and spark new research interests and developments in this field.